\documentclass{vgtc}                          

\ifpdf
  \pdfoutput=1\relax                   
  \usepackage{graphicx}                
  \DeclareGraphicsExtensions{.pdf,.png,.jpg,.jpeg} 
\else
  \usepackage{graphicx}                
  \DeclareGraphicsExtensions{.eps}     
\fi%

\graphicspath{{figures/}{pictures/}{images/}{./}} 

\usepackage{microtype}                 
\PassOptionsToPackage{warn}{textcomp}  
\usepackage{textcomp}                  
\usepackage{mathptmx}                  
\usepackage{times}                     
\usepackage{cite}                      
\usepackage{tabu}                      
\usepackage{booktabs}                  
\usepackage{enumitem}

\usepackage{bbding} 
\usepackage{amsmath} 
\usepackage{adjustbox} 
\usepackage{multirow} 

\usepackage[final]{changes}  
\usepackage{ragged2e} 
\usepackage{caption}
 
\usepackage{color}
\definecolor{green_2}{rgb}{0.1,0.8,0.3}
\definecolor{mauve}{rgb}{0.48,0,0.72}

\onlineid{1712}

\vgtccategory{Research}

\title{Exploring the Role of Expected Collision Feedback \\ in Crowded
Virtual Environments}

\author{Haoran Yun\thanks{e-mail:haoran.yun@upc.edu} %
\and Jose Luis Ponton\thanks{e-mail:jose.luis.ponton@upc.edu} %
\and Alejandro Beacco\thanks{e-mail:alejandro.beacco@upc.edu} %
\and Carlos Andujar\thanks{e-mail:carlos.andujar@upc.edu} %
\and Nuria Pelechano\thanks{e-mail:nuria.pelechano@upc.edu}}


\vgtcauthor{Haoran Yun, Jose Luis Ponton, Alejandro Beacco, Carlos Andujar, Nuria Pelechano}

\vgtcinsertpkg

\affiliation{\scriptsize Universitat Politècnica de Catalunya, Spain}

\teaser{
  \justifying
  \includegraphics[width=\linewidth, alt={(a) Three types of collision feedback tested in the study (b) Adaption scenario (c) Corridor scenario (d) Crossing scenario. Details of the factors and scenarios can be found in Section 3.} ]{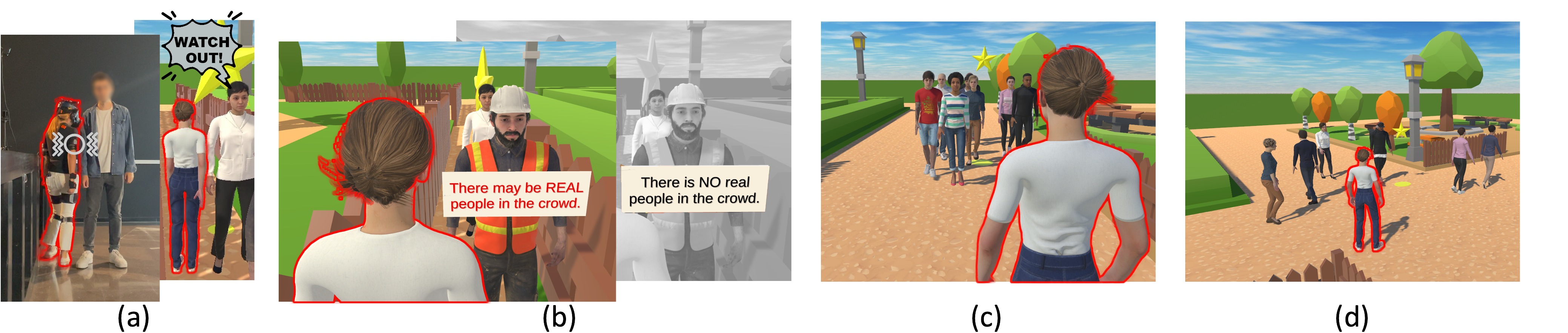}
   \captionlistentry{}
  \label{fig:teaser}
} {We explore the effect of collision feedback expectancy in users who move through groups of virtual agents. We tested nine conditions, differing in the combination of types of collision feedback, shown in (a), users \textit{could expect} when colliding with a virtual agent: an audible complaint, a vibration provided by an arm-mounted vibrotactile device, and a physical bump with a real person. We designed three different scenarios: a narrow corridor, shown in (b), where collisions with virtual agents were inevitable, a corridor, shown in (c),  where agent movement was easily predictable and collisions could be avoided by taking a detour around the group, and a street crossing, shown in (d), where virtual agents moved in two orthogonal directions. The avatar of the participant is highlighted in red.} 

\abstract{
An increasing number of virtual reality applications require environments that emulate real-world conditions. These environments often involve dynamic virtual humans showing realistic behaviors. Understanding user perception and navigation among these virtual agents is key for designing realistic and effective environments featuring groups of virtual humans. While collision risk significantly influences human locomotion in the real world, this risk is largely absent in virtual settings. 
This paper studies the impact of the expected collision feedback on user perception and interaction with virtual crowds. We examine the effectiveness of commonly used collision feedback techniques (auditory cues and tactile vibrations) as well as inducing participants to expect that a physical bump with a real person might occur, as if some \replaced{virtual humans}{avatars} actually correspond to real persons embodied into them and sharing the same physical space.
Our results indicate that the expected collision feedback significantly influences both participant behavior\textemdash encompassing global navigation and local movements\textemdash and subjective perceptions of presence and copresence. Specifically, the introduction of a perceived risk of actual collision was found to significantly impact global navigation strategies and increase the sense of presence. Auditory cues had a similar effect on global navigation and additionally enhanced the sense of copresence. In contrast, vibrotactile feedback was primarily effective in influencing local movements.

}

\CCScatlist{
  \CCScatTwelve{Computing methodologies}{Computer graphics}{Graphics systems and interfaces}{Virtual reality, Perception};
  \CCScatTwelve{Human-centered computing}{Human computer interaction (HCI)}{HCI design and evaluation methods}{User studies};
}

\begin{document}



\firstsection{Introduction}
\label{sec:intro}

\maketitle

In computer applications such as building evacuation planning, training, and video games, crowd simulation is becoming increasingly important, as it allows hundreds or thousands of agents to navigate through the virtual environment. 
Furthermore, a number of applications require the user to be immersed in such virtual environments. 
By using a VR headset, the user can be embodied into a virtual avatar \cite{Ponton:2023}, gaining an illusion of body ownership over it, feeling present (sense of presence) 
in the virtual world and having the experience of being part of the virtual crowd. 

Presence can be seen as the addition of place illusion (PI) and plausibility (Psi) \cite{slater2009,slater2022}. 
PI is the illusion of being in the place depicted by the VR.
Psi is the illusion that the virtual situation and events are really happening. 
While PI can be achieved in VR through natural sensorimotor contingencies integrated by the hardware on sensory outputs \cite{oregannoe2001},
Psi is a far more complex and interesting aspect to study since any unexpected event can produce a break in presence or a failure of expectations in the participant. 
Similarly, copresence refers to the extent to which a participant has the illusion of being there with other participants \cite{durlach2000}. This can also be achieved through sensorimotor contingencies that give the participant the illusion of being in the same space as the other humans, whether real or virtual, as it happens with PI. 
To achieve Psi, it is also necessary that the virtual characters respond to the user when the user interacts with them. The most difficult requirement for copresence is meeting the user's expectations, which largely depend on the context (for example, people behave differently in a park vs.\ in a bus) and appearance (cartoon characters will imply different expectations than realistic ones) \cite{Zibrek2018}. 
When copresence is achieved, we can expect users to show similar behaviors to those in reality, such as maintaining the same proxemics \cite{Bailenson2003}.

Regarding virtual crowds, the breaks in presence or failures of expectations can be due to different factors such as visual or animation artifacts (e.g., foot sliding \cite{pelechano2011}), detecting \replaced{virtual character}{avatar} clones \cite{mcDonnell2008}, strange behaviors, or lack of interaction. 
If virtual agents do not respond to the user's actions, the user might quickly notice this and cause a break in presence.  

Conversely, we should expect that some actions or events happening in the virtual environment could also affect the user's behavior. 
In our case, we wonder how a virtual crowd may affect the user's behavior, and more specifically, how collisions with virtual agents may influence the user depending on different types of collision feedback users might expect: an audible complaint (audio), a vibrotactile feedback through an arm-mounted device, or a physical bump with a real person embodied in an avatar. 
The first two feedback modalities can be easily implemented with commodity VR hardware, so all collisions with a virtual agent could trigger any combination of these feedback modes. In contrast, providing realistic physical bump feedback for all collisions is impractical, as this would imply that all avatars correspond to physical users embodied in them. Instead, we designed the experiment so that users are induced (depending on the condition) to expect a physical bump with a real person, even if such bumps occur only in an initial adaptation scenario. 
Similar to the Rubber Hand Illusion \cite{botvinick1998}, we physically touch the participants in synchrony with the virtual collision to induce such belief.
We assume this adaptation will also induce a collision feedback expectancy from the point of view of the user. 
Therefore, our study involves mainly exploring the role of this expected collision feedback in crowded virtual environments. Our specific research questions are:


\begin{itemize}

\item \textbf{RQ1:} Does collision feedback expectancy (users induced to expect a combination of audio, vibrotactile, or physical bump feedback) enhance the \textbf{sense of presence and/or copresence} in tasks involving moving across a group of virtual agents?
    \item \textbf{RQ2:} Does collision feedback expectancy influence the \textbf{behavioral response} of individuals in the tasks above? 
\end{itemize}



Our results show that expected collision feedback plays an important role in shaping participants’ behavior and subjective perceptions towards virtual crowds. In the absence of it, people might tend to ignore other agents. However, believing there could be physical collisions with real people can enhance presence. Also, copresence with the virtual crowds can be enhanced by adding audio feedback. Similarly, global behaviors such as navigation can also be significantly influenced: participants will show a tendency to maintain extended distances from crowd agents and take more time to complete tasks. Contrarily, vibrations can affect local behaviors such as adding more torso motion to actively avoid collisions.

\section{Related Work}
\label{sec:related_work}

The existing body of literature on crowd behavior in VR environments encompasses several areas, including crowd perception, realistic crowd simulation, and collision feedback mechanisms. For the purpose of this paper, we divide related work into three broad categories: studies related to crowd perception in VR, \replaced{proxemics and copresence}{social presence}, and providing collision feedback in virtual crowds.

\subsection{Virtual Crowds}

An essential factor in crowd simulation is the accurate representation and perception of crowds. Olivier \textit{et al.} \cite{olivier:2014} establish the efficacy of VR as a tool for the study of crowds. 
Subsequent studies have delved into various crowd attributes that could influence users' perception in VR. Factors like the visual appearance of the crowd \cite{mousas2021} and the realism and diversity of animations \cite{molina:2021} have been shown to enhance the perceived authenticity of the virtual crowd, affecting how users navigate around other agents. However, it is worth noting that an exhaustive set of animations is not essential; minimal motion variation is adequate for maintaining the perception of crowd heterogeneity, especially in scenarios involving large crowds \cite{adili:2021}. Additionally, the presence of crowds has been shown to influence the decision-making processes of users in a VR environment \cite{rios:2020}.

Raimbaud \textit{et al.} \cite{raimbaud:2023} investigate the impact of agents' gaze on users, highlighting that gaze direction may affect those with social anxiety but not necessarily impact their locomotion. In this line, some works have focused on studying how the emotional state of a virtual crowd may affect the way people navigate and interact  \cite{patotskaya:2023, volonte2020}. Adding basic social interaction can also increase crowd realism and user presence \cite{kyriakou:2017}. Similarly, responsive virtual crowd behaviors increase the feeling of presence \cite{kyriakou:2018}.

Evidence of the necessity for realistic crowd behavior in VR comes from works that simulate real-world scenarios. Arias \textit{et al.} \cite{arias:2019} reconstructed the Love Parade disaster in VR to evaluate various disaster prevention strategies. Zhao \textit{et al.} \cite{zhao:2020} used VR to test different fire evacuation systems for a building under construction, highlighting the cost-effectiveness of VR in such studies.

\subsection{Proxemics and copresence}
User's locomotion behavior when moving around other avatars in immersive VR can vary depending on several factors, such as the quality of the animations and the fear of a real collision happening. For example, the work by Ríos \textit{et al.} \cite{rios2018} showed that when users walk around another virtual human while sharing the same virtual space (thus, there is the possibility of a real collision), proxemics could be very similarly to reality, as long as the animations of the avatars were continuous. Previous work observed larger proxemics or clearance not being respected when animation artifacts were present, thus leading to virtual collisions when users knew the other user was located remotely \cite{podkosova2018mutual}. \deleted{It has also been observed that users tend to walk slower in VR than in the real-world counterpart.} \added{Trivedi \textit{et al.} \cite{trivedi:2023} found that a high avoidance radius in the simulated crowd agents led to longer paths and lower perceived realism, drastically reducing user experience.}
Some researchers have investigated the impact of the avatar's appearance on users and observed that photorealism can increase plausibility and presence, but that it does not seem to affect copresence \cite{McDonnell2012}. In fact, their study showed that what increases copresence is the user-avatars interaction, for example, if the avatars gaze at the user.


\subsection{Collision Feedback in VR}

Understanding collision feedback in virtual environments is critical for simulating realistic crowd behaviors. The work by Olivier \textit{et al.} \cite{olivier2012} highlights that pedestrians only adjust their movements if they perceive a future risk of collision. This observation is particularly relevant to our study, as it suggests the necessity for exploring how expected collision feedback might affect collision avoidance behavior in VR crowds.

A common approach for simulating collision feedback is using haptics. Krogmeier \textit{et al.} \cite{krogmeier2019a} used a haptic vest equipped with 70 haptic points and showed that haptic feedback enhances presence and embodiment. On the other hand, Berton \textit{et al.} \cite{berton:2022} reported that haptic feedback improved collision avoidance but did not significantly affect the sense of presence. Similarly, Koilias \textit{et al.} \cite{koilias:2020} demonstrated that users' behavior was affected when haptics was enabled; however, they failed to distinguish between random and accurate haptic feedback. Similarly, Krum \textit{et al.} \cite{krum2018} studied how priming haptic rendering influenced social measures and found that it impacted subjective social experiences but not proxemics. \deleted{Trivedi \textit{et al.} found that a high avoidance radius in the simulated crowd agents led to longer paths and lower perceived realism, drastically reducing user experience.} \added{Different types of haptic feedback also led to different levels of emotional response towards virtual crowds. More recently, Venkatesan \textit{et al.} \cite{Venkatesan2023} observed that users reported a higher sense of realism and increased nervousness when exposed to haptic stimuli, as opposed to solely tactile stimuli or no feedback at all.}

Audio cues have mainly been given as a sound informing the participant about a collision, for example, while navigating using different locomotion techniques \cite{suma2007}. Blom \textit{et al.} introduced the soundfloor, an audio-haptic interface for providing virtual collision feedback with the ground, but their study showed no effect of such feedback on performance \cite{blom2012}. On a different kind of task, insertion tasks (manipulating a 3D object to get through one or more virtual apertures), Lecuyer \textit{et al.} also studied how adding haptic, visual, and audio feedback affected performance \cite{lecuyer2002}. While none of the added cues improved the task completion time, it seemed that participants moved less when colliding and having additional feedback, as if paying more attention to the possible collisions, and therefore taking more time to complete their task. 

Studying social VR, Reinhard and Wolf explored the usability of applications involving virtual avatars and how difficult some tasks can be when they block paths or occlude the participant's view \cite{reinhardt2020}. Although opposite from our goal, they propose a solution without collision but where the participant is allowed to walk through the other avatars. They then explore multimodal feedback related to that, such as hearing the heartbeat of the avatar you are walking through. In their results, they find that presence significantly increases when multimodal feedback is provided, referring to our real-world experience when bumping into someone, but their study was limited to a two-avatar configuration. 

\added{Previous research has primarily explored the impact of single or multimodal collision feedback, such as audio and haptics, on users’ interactions and perceptions within virtual crowds. Yet, there is still a lack of studies investigating the influence on users’ behavior and perception when they expect potential collisions with real individuals in a virtual crowd setting. This scenario presents a difference to the effects evoked by commonly employed collision feedback techniques. Furthermore, our research aims to bridge this gap by examining the combined effects of these feedback modalities on user experience.}


\section{User study: overall design}
\label{sec:design}

This section outlines the main decisions regarding the user study: the 
independent variables we considered, and how we implemented them in the virtual reality application. The scenarios and the activities participants had to undertake within the scenarios are also detailed, along with other relevant aspects concerning the appearance and behavior of the virtual agents. 

\subsection{Factors: Collision Feedback Expectancy}
\label{sec:design:factors}

In the real world, collision risk with other individuals is a critical factor that explains crowd behavior. This risk leads individuals to move cautiously and exhibit distinct behavioral patterns compared to isolated walking. To simulate this aspect in a VR environment, we believe it is necessary to induce in users a certain expectancy of collision feedback, influencing their actions and locomotion choices to align with real-world behavior. Existing methods often use auditory cues or tactile feedback to achieve this outcome. The ultimate goal of these methods is to establish a collision feedback expectancy, encouraging users not to disregard virtual characters but to perceive them as real entities sharing the same physical space (copresent) and thus push them to avoid collisions.

A basic crowd simulation in VR entails rendering diverse human avatars moving along specified paths. Although this alone may facilitate a sense of presence, we believe that it does not prevent users from disregarding the crowd and going through virtual characters. We hypothesize that introducing collision feedback expectancy will increase the sense of presence and copresence, and lead users to move more cautiously.

The following are the specific factors related to collision feedback expectancy examined in this study:

\begin{enumerate}[label=(\alph*)]
    \item \textit{AUDIO}: Positional audio (a clearly audible complaint) upon collision with a virtual character.
    \item \textit{VIBR}: Arm-specific vibration upon collision with a virtual character.
    \item \textit{COLBLF}: Inducing the feeling or belief that a collision risk with an actual human exists, showing a virtual panel and actually having a gentle physical collision during an adaptation phase.
\end{enumerate}

When the \textit{AUDIO} is enabled, a collision triggers positional audio cues emitted from the location of the involved virtual character. To maintain consistency while minimizing auditory repetition, three gender-specific audio samples were chosen, resulting in a total of six distinct audio cues. Examples of these samples include \emph{Ouch!}, \emph{Careful!}, and \emph{Watch Out!}.

For \textit{VIBR}, participants are equipped with two vibration devices on their upper arm exclusively activated upon collision with virtual avatars. The vibration is selectively initiated in the arm closest to the estimated collision point. Specifically, if a collision with an avatar occurs on the participant's left side, the left armband will be activated, and correspondingly, collisions on the right arm trigger vibrations in the right armband. 

Regarding the \textit{COLBLF} factor, we hypothesize that by nurturing the belief that genuine collisions can occur, users would exhibit collision avoidance behaviors more consistent with real-life crowds. To strengthen this belief, during an adaptation phase, users bump into a real person's arm when the \textit{COLBLF} setting is enabled. 

\subsection{Simulation design}
\label{sec:design:crowd}
\subsubsection{Main scenarios}
We designed two main scenarios: a single corridor (\textit{Corridor}) and a street crossing (\textit{Crossing}). \replaced{The corridor had a 4.5\,m$\times$7.5\,m walkable area, both physically and virtually, for participants walking in naturally.}{The corridor had a width of 4.5\,m and a total simulation length of 10\,m. However, the length of the physical space was 7.5\,m; thus, the user had to move within an area of 33.75\,m$^2$ to make navigation possible through natural walking. } Similarly, for the Crossing scenario, users walked in the intersection of two corridors, \replaced{which had a 4.5\,m$\times$4.5\,m walkable area}{which were approximately 4.5\,m wide, yielding a total area of 20.25\,m$^2$}. \added{Virtual humans were spawned slightly outside the walkable area to allow users more time to observe the agent's trajectories.}

\subsubsection{Crowd trajectory} The crowd simulation was implemented using the RVO2-Unity package, which incorporates the Optimal Reciprocal Collision Avoidance (ORCA) algorithm. The maximum speed was set to 0.75\,m/s \added{which is around 62.5\,\% of normal walking speed 1.2\,m/s, according to the U.S. Manual on Uniform Traffic Control Devices. When navigating around other virtual humans, people tend to walk slower \cite{rios2018}, especially if haptic feedback is provided then the walking speed drops to 0.4\,m/s approximately \cite{berton:2022}. Thus, a relatively low speed was chosen in our study to allow participants to have enough time to plan and navigate through the crowd.} 
With these settings, along with the initial position, rotation, and target position, RVO2 generated a 2D global trajectory for each character.

\begin{figure}[tb]
 \centering
 \includegraphics[width=1\columnwidth, alt = {Top: Corridor scenario showing the direct path with a continuous line and the detour path with a dotted line. Bottom: Crossing scenario showing a busy street crossing with a virtual crowd. }]{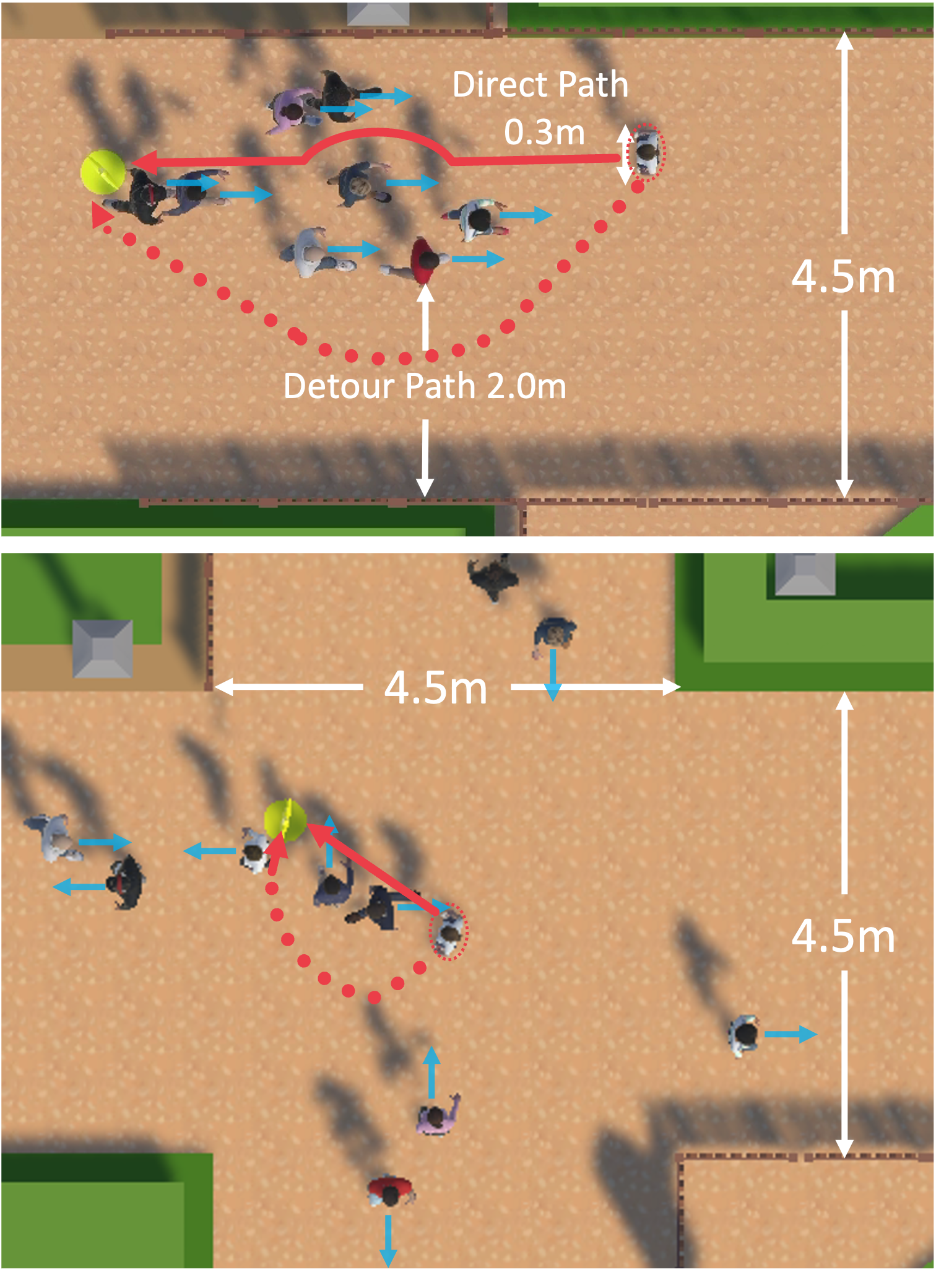}
 \vspace*{-6mm}
 \caption{Corridor scenario showing the direct path with a continuous line and the detour path with a dotted line (\textit{top}). Crossing scenario (\textit{bottom}).}
 \label{fig:tasks}
\end{figure}

In the Corridor scenario, we confined the spawn and target positions of the characters to ensure that participants had at least two path options across the moving agents. More precisely, the crowd design incorporated two distinct pathways: a 0.3-meter-wide \textit{direct path} placed closer to the user, and a wider 2-meter-wide \textit{detour path} that required a larger detour to avoid the incoming crowd, as shown in Figure~\ref{fig:tasks} top. The direct path served as a more direct route to the goal but increased the likelihood of collision with the crowd agents. Conversely, the detour path offered a safer but slower route. Our hypothesis posits that in the absence of collision-related feedback, participants would be more likely to opt for the \textit{direct path}. 

In the Crossing scenario, two orthogonal directions were included resulting in a crossing between two corridors. Agents traversed from one end of the crossing to the other and reverse\added{d} direction upon reaching the end. To achieve this behavior, four spawn positions were located at four ends of the crossing. Each agent was assigned a specific target. Once reaching the target, another target in the opposite direction was created. This scenario is intended to more closely mimic a real crowd environment in which individuals move in multiple directions and have different objectives. Virtual agents were programmed to avoid colliding with each other and the participant \added{only in the Crossing scenario}.

\subsubsection{Adaptation Scenario}
\label{sec:design:training}

Besides the two main scenarios described above, we also designed an adaptation scenario to serve during an initial phase conceived to accommodate the participant to the current experimental condition. 
In this scenario, the participant was placed in a narrow corridor alongside one notice board and two virtual \replaced{humans}{avatars} standing with an idle animation. The task was to reach a designated target position, experiencing a collision with both virtual characters. The feedback associated with the given condition\textemdash a combination of \textit{AUDIO}, \textit{VIBR}, or \textit{COLBLF}\textemdash was applied when these collisions occurred.  

When the \textit{COLBLF} factor was enabled, collision belief was induced through a visual warning and physical bumps in this adaptation scenario. The visual warning, ``There may be REAL people in the crowd'' in red, was displayed on the notice board, as shown in Figure~\ref{fig:teaser} (b). The physical bump was facilitated by a confederate standing at the location of one of the two \replaced{virtual characters}{avatars} in the real world. As participants passed by this position, the experimenter physically bumped into them softly. We ensured that the physical contact was gentle and safe. 

In contrast, when the \textit{COLBLF} factor was disabled, the notice board showed ``There are NO real people in the crowd.'' and the confederate was not standing in the \replaced{virtual characters}{avatar}'s position. 

It is noteworthy that a physical collision between the experimenter and participants occurred only once during the adaptation phase for each condition with \textit{COLBLF} enabled.

\subsubsection{Agents' appearance} We selected a set of 10 character models from Adobe Mixamo \cite{mixamo2023}, comprising \replaced{5}{6} male and \replaced{5}{6} female \replaced{characters}{avatars}. The distribution of models across scenarios was as follows: 8 for Corridor and 10 for Crossing. We ensured an even representation of male and female characters in each scenario and no clones in the crowd. 
In order to achieve a sense of copresence while avoiding the uncanny valley, we opted for a cartoonish visual style for our virtual crowd \replaced{characters}{avatars}. This stylistic choice was consistently applied to the virtual environment as well. Our decision is supported by prior research indicating that although photorealistic rendering can enhance the sense of presence, non-photorealistic environments are also effective in achieving high levels of user presence \cite{McDonnell2012}.

\subsubsection{Agents' animation} We used a data-driven method called Motion Matching to synthesize character motions \cite{Ponton:2022b}. This method uses a set of features, including positions and velocities of the feet, the velocity of the hips, and future trajectory directions and positions. Then this information is used as query vectors to search over an animation database for the best matches. The animations included in the dataset were captured using the Xsens Awinda system, with two actors (one female, one male) walking at varying speeds: slow, normal, and fast, as well as walking in different directions like forward, backward, turning, and in-place rotations. In contrast to using single or multiple walking cycles, this data-driven method provides characters with more responsive and realistic walking motions, avoiding potential motion artifacts like foot sliding (see accompanying video). 

\section{User Study: design details}
\label{sec:user_study}

\subsection{Apparatus}
\label{sec:user_study:apparatus}

The experiment was conducted in an isolated laboratory room with 7.5m$\times$4.5m free walking space. The VR application was developed in Unity 2021.3.12f1 and ran on a PC equipped with an Intel Core i7-12700, NVIDIA GeForce RTX 3070 Ti, 32\,GB RAM. As a VR headset, we used an HTC Vive Pro, which has a resolution of 1440 × 1600 pixels per eye, $110^{\circ}$ field of view, 90\,Hz refresh rate, and 3D spatial sound (using the HMD headphones). To provide users with a safe and spacious walking area, a Vive Wireless adapter was used instead of a wired connection. Four SteamVR Base Stations~2.0 were placed at each corner of the room to minimize potential occlusions. A wireless IMU-based motion capture system with 17 sensors, Xsens Awinda, was deployed to give users full-body animation in VR. Myo armbands, gesture control armbands from Thalmic Labs, generated a 1.5-second vibration for each collision, if \textit{VIBR} factor was enabled. 

\subsection{\replaced{Study Design}{ Factorial design}}
\label{sec:user_study:design}
We adopted a within-subjects design, where each of the three factors had two levels, \emph{enabled} and \emph{disabled}, resulting in eight conditions. Participants sequentially completed the Adaptation, Corridor, and Crossing scenarios for each condition.
An embodiment-inducing phase, where the participant could get used to the virtual avatar, was presented at the beginning of the experiment. 
Additionally, a pretest condition was also conducted with all three factors set to \emph{disabled} prior to the main experiment. This pretest \cite{bonate2000analysis} allowed us to evaluate the sensitization effects, such as learning or after effects of the main factors. To counterbalance the order of conditions between participants, we used an $8 \times 8$ Balanced Latin Square. Crowd configurations and target positions remained consistent across different conditions and participants.

\subsection{Procedure}
\label{sec:user_study:Procedure}
A total of 16 participants took part in the experiment (6 female, 10 male, aged 22-39, $\mathrm{\bar{x}} = 26.7$, $\sigma = 5.0$). On arrival, participants were briefed via a document detailing the tasks they would perform, followed by completing a consent form and a demographic questionnaire. After the briefing, they were equipped with the two Myo armbands (one on each upper arm) and the Xsens motion capture suit. After calibrating the Xsens system, participants were equipped with the VR headset.

Next, participants underwent an embodiment phase, immersed in a virtual room with a mirror, allowing them to move and observe their virtual body freely for about three to four minutes. This phase facilitated inducing the sense of embodiment with the virtual body. Afterward, we started with the pretest, followed by the eight conditions of the main experiment, randomized as described above. 

At the beginning of each condition, participants started in the middle of a narrow corridor in the Adaptation scenario, embodied in a gender-matched avatar. They were instructed to read the notice board first, then proceed toward a target location marked by a star-shaped object. Post-adaptation, participants transitioned to the Corridor and the Crossing scenarios.

In both scenarios, participants were instructed to walk toward a clearly visible target. Once reached, a new target and a new virtual crowd were generated. No instructions were given about how participants should behave in terms of trying to avoid collisions.

For the corridor scenario, each new target was created at the other end of the corridor, so that in each trial, participants had to walk through a new group of agents walking against them. The simulated group performed no collision avoidance, thus leading to inevitable collisions if the participant remained stationary. The goal of this simulated behavior was to study participants' route choices between the two visible paths left by the crowd formation (the \textit{Direct Path} or the \textit{Detour Path}). Regarding the crossing scenario, participants were positioned at the intersection between the corridors and were also asked to reach the designated target positions. In this scenario, simulated characters would perform collision avoidance against each other and participants.

At the end of each condition, they were required to complete a questionnaire within the VR environment (see Table~\ref{tab:questions}), and then they could take a short break. After finishing all the conditions, they were asked to finish a short survey of open-ended questions (see Table~\ref{tab:open_questions}).



\subsection{Measures}
\label{sec:design:measures}

\begin{table}
\scriptsize
    \centering
    \begin{tabu}{lX}
    \toprule 
    \multicolumn{2}{l}{\textit{\textbf{Presence}}} \\ 
    \multicolumn{2}{l}{\textit{\small{\textit{(Reversed P1 + P2 + Reversed P3 + P4) / 4}}}} \\
    \cmidrule(lr){1-2}
    \textbf{P1} & I did not feel present in the virtual space. \\
    \textbf{P2} & I felt present in the crowd. \\
    \textbf{P3} & I still paid attention to the real environment.  \\
    \textbf{P4} & Somehow, I felt that the virtual world surrounded me. \\
    \cmidrule(lr){1-2}
    \multicolumn{2}{l}{\textit{\textbf{Copresence}}} \\
    \multicolumn{2}{l}{\small{\textit{(CP1 + CP2 + Reversed CP3 + CP4 + Reversed CP5) / 5}}} \\
    \cmidrule(lr){1-2}
    \textbf{CP1} & I perceived that I was in the presence of other people with me. \\
    \textbf{CP2} & I felt that people were aware of my presence. \\
    \textbf{CP3} & The thought that people were not real crossed my mind often. \\
    \textbf{CP4} & The people appeared to be alive to me. \\
    \textbf{CP5} & I perceived people as being only a computerized image, not as real people. \\
    \bottomrule \addlinespace[0.05cm]
    \end{tabu}
     \caption{Questionnaire content. The scores are on a 5-Likert scale (1~= strongly disagree, 5 = strongly agree).}
    \label{tab:questions}

\end{table}

\subsubsection{Objective metrics} 
\label{subsubsec:objective}
\begin{figure}[tb]
 \centering
 \includegraphics[width=1\columnwidth, alt={Visual illustration of Clearance and Torso measures. A detailed description of the measures can be found in Section 4.4.1.}]{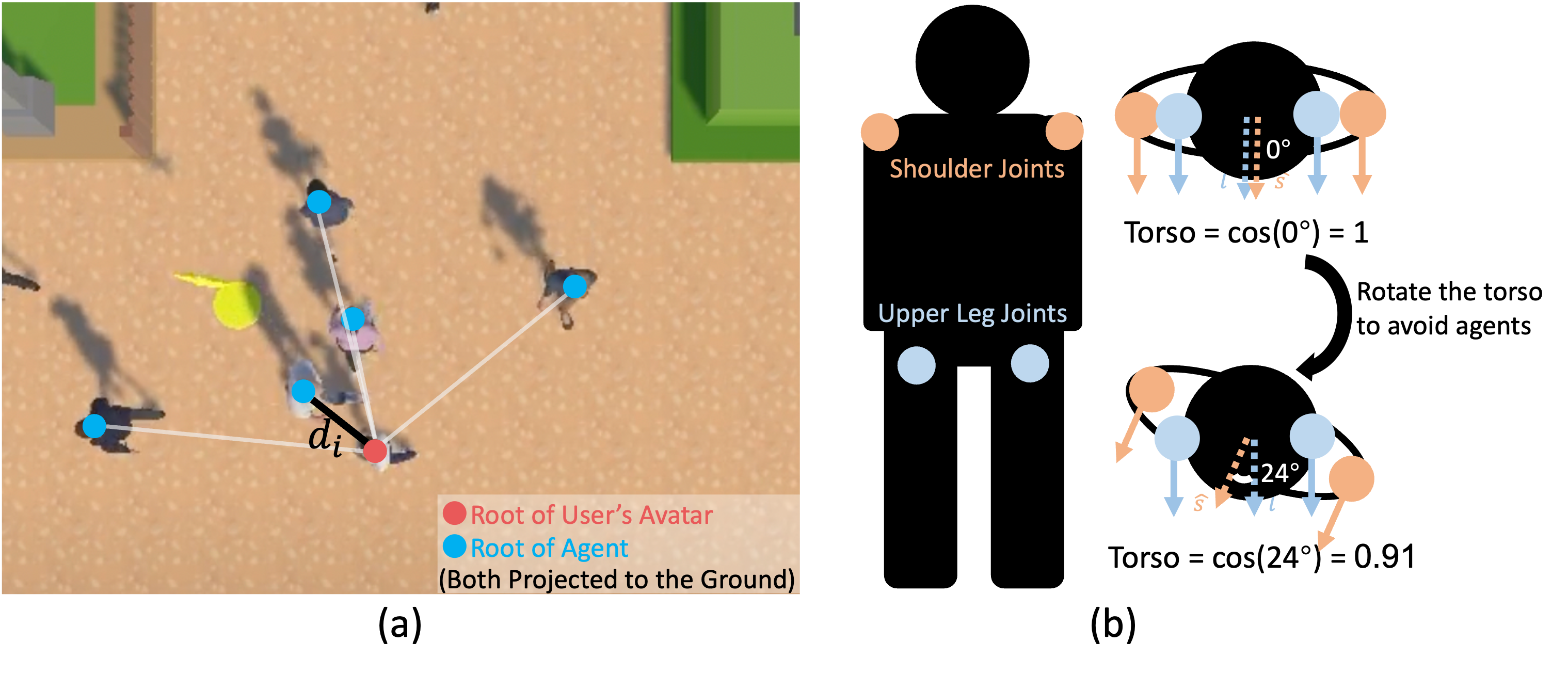}
 \vspace*{-6mm}
 \caption{\added{Visual illustration of \textit{Clearance} and \textit{Torso} measures. (a) The distances between the user avatar's root position and every agent's root position are calculated at timestamp $i$ (projected onto the ground); $d_i$ is the shortest distance at this timestamp.} \added{ (b) \textit{Torso} refers to the cosine of the angle, $\mathbf{\alpha}$, between the torso forward vector (in light salmon), $\mathbf{\hat{s}}$, and the hips forward vector (in light blue), $\mathbf{\hat{l}}$. The torso and hips forward vectors are computed from the forward vectors of shoulders and upper legs respectively. } \added{The \textit{Torso} value changes when a user rotates the torso to avoid nearby agents.}}
 \label{fig:torso_and_clearance}
\end{figure}

To objectively evaluate participants' behavior, we recorded diverse data from the simulation, including the trajectories of participants and agents on the ground plane within the virtual environment, the completion time for each task, and the number of collisions with virtual agents. Concurrently, we captured body motion using the Xsens motion capture system for further postural analysis \added{(e.g., torso rotation)}. The collected data was further processed to obtain more intuitive metrics as follows:
\begin{itemize}
    \item \textit{Collisions}  = $\sum_{i = 1}^{N} c_{i}$, where $N$ is the number of trials, and $c_{i}$ refers to the number of detected contacts between the participant's avatar and the agents, for each trial. 
    
    \item \textit{Time} = $\frac{1}{N}\sum_{i = 1}^{N} t_{i}$, where $N$ is the number of trials, and $t_{i}$ refers to completion time for each trial. 
    
    \item $\textit{Clearance} = \frac{1}{0.1T} \sum_{i = 1}^{0.1T} d_i$, where $T$ is the number of timestamps (with a rate of 50 per second) captured during the simulation. Let $\{d_0, d_1, \ldots, d_T\}$ be the array of clearances per timestamp, sorted in ascending order. Here, $d_i$ represents the distance \replaced{between the root position of user's avatar projected onto the ground and the root of }{to} the nearest agent \added{projected onto the ground} at each timestamp\added{, as depicted in Figure~\ref{fig:torso_and_clearance} (a)}. As the focus is on moments of minimum clearance and considering that the user may often be distant from the crowd, the average distance is calculated using only the lowest $10\,\%$ of the samples.
    

    \item \textit{Torso} = $\mathbf{\hat{l}} \cdot \mathbf{\hat{s}}$, \added{which provides the cosine of the angle, $\mathbf{\alpha}$, between the torso forward vector, $\mathbf{\hat{s}}$, and the hips forward vector, $\mathbf{\hat{l}}$. The torso forward vector 
    $\mathbf{\hat{s}}$ is computed as the average of both shoulder joints forward vectors. Similarly, $\mathbf{\hat{l}}$ is computed from both upper leg joints forward vectors (this information is extracted from the motion capture suit), see Figure~\ref{fig:torso_and_clearance} (b). Similarly to $\textit{Clearance}$, we order this value for all timestamps and consider only the lowest $10\,\%$ of the samples.
   The cosine provides an intuitive estimate of the user intention to reduce the effective width, $\mathbf{eW}$, which would be the result of projecting the vector between the shoulder joints onto a direction perpendicular to the displacement, and it can be calculated as $\mathbf{eW} = cos(\alpha) \cdot \mathbf{sW}$, with $\mathbf{sW}$ being the shoulder width. }\deleted{ where $\mathbf{\hat{l}}$ is the average of both upper leg joints forward vectors, and $\mathbf{\hat{s}}$ is similarly defined for both shoulder joints. Similarly to $\textit{Clearance}$, we order this value for all timestamps and consider only the lowest $10\,\%$ of the samples. We capture this information with the pose data extracted from the motion capture suit. In our setting, we assume that significant torso rotations indicate participants' intent to avoid collisions with a nearby avatar.}

    \item  \textit{Path} = 
    $\begin{cases}
      0 & \text{if the participant chose a direct path} \\
      1 & \text{if the participant chose a detour path}
    \end{cases}$ \\
    This binary value is defined for each trial of the Corridor scenario. A manual labeling process was carried out post-experiment to classify participants' path choice\textemdash direct or detour\textemdash based on the recorded trajectories.
    
\end{itemize}
Due to the distinct characteristics of the Corridor and Crossing scenarios, specifically in terms of crowd simulation, we opted to analyze certain metrics separately for each scenario. Specifically, we examined $\mathit{Collisions_{corridor}}$, $\mathit{Collisions_{crossing}}$, 
$\mathit{Time_{corridor}}$, $\mathit{Time_{crossing}}$, $\mathit{Clearance_{corridor}}$ and $\mathit{Clearance_{crossing}}$.
\textit{Torso} was analyzed once per condition regardless of the scenario, 
and \textit{Path} was measured only for the Corridor.

\added{
Furthermore, our analysis focuses on the interactions between users and nearby agents across all tasks. Notably, users often remain idle, awaiting subsequent tasks, or maintain a large distance from agents. Our initial strategy involved setting thresholds to omit irrelevant clearance values\textemdash particularly when users are significantly distant from agents. However, this approach proved insufficient, as certain participants consistently maintained clearance values above our predetermined threshold, thereby yielding no comparative data. Consequently, we adopted a more universally applicable method suitable for diverse tasks. This method involves organizing the data and selectively extracting the most pertinent percentage. Specifically, for the \textit{Clearance} task, we extract the minimal distances to agents, and for tasks involving torso rotation, we focus on the maximal angle deviations. Upon analyzing the data distribution, we determined that the chosen $10\,\%$ subset effectively balances comprehensiveness and relevance for our experimental needs. 
}

\added{
Additionally, it is important to highlight that we conducted the statistical analysis both with and without the implemented cutoffs, encompassing all \textit{Clearance} distances and \textit{Torso} values. The results remained consistent in both scenarios. 
}

\subsubsection{Subjective responses} 
\label{subsubsec:subjective}

We also collected subjective ratings with a questionnaire at the end of each condition. The questionnaire included nine items, four addressing \textit{Presence} and five addressing \textit{Copresence}, as detailed in Table~\ref{tab:questions}. Questions were adapted from the I-Group Presence Questionnaire (IPQ) \cite{Schubert:2001, schubert2003sense} and the copresence questionnaire by Bailenson \textit{et al.} \cite{Bailenson2003}. 

To further analyze participants' experiences, we posed several open-ended questions concerning the overall experiment and each factor, as shown in Table~\ref{tab:open_questions}. We encouraged participants to write down their opinions without limits in a text file. Responses were categorized into positive and negative sentiments using \emph{sentimentr} R package \cite{rinker:2021}.

\begin{table}
\scriptsize
    \centering
    \begin{tabu}{lXccc}
    \toprule
    \multicolumn{2}{l}{Open-Ended Question} & $n$ & $\mathrm{\bar{x}}$ & $\sigma$ \\
    \cmidrule(lr){1-2} \cmidrule(lr){3-5}
    \textbf{Q1} & What can you tell us about the overall experience? & 886 & 0.09 & 0.36 \\ \addlinespace[0.1cm] 
    \textbf{Q2} & How did you feel when people reacted to collisions with audio? & 431 & 0.09 & 0.30 \\ \addlinespace[0.1cm] 
    \textbf{Q3} & How did you feel when you felt vibrations in your arms? & 471 & -0.09 & 0.32 \\ \addlinespace[0.1cm] 
    \textbf{Q4} & Believing you may collide with a real person, did it make you behave more like in real life? & 433 & 0.21 & 0.36 \\ 
    \bottomrule\addlinespace[0.05cm]
    \end{tabu}
     \caption{Post-experiment open-ended questions. We report the results of the sentiment analysis (i.e., the positivity of the answers) performed with the \emph{sentimentr} R package \cite{rinker:2021}; $n$ is the number of words analyzed, $\mathrm{\bar{x}}$ is the average sentiment, and $\sigma$ is the standard deviation.}
    \label{tab:open_questions}
\end{table}

\section{Results}
\label{sec:results}

\begin{figure*}
    \centering
    \includegraphics[width=1\linewidth, alt = {Box plots for all objective and subjective measurements, except Path.}]{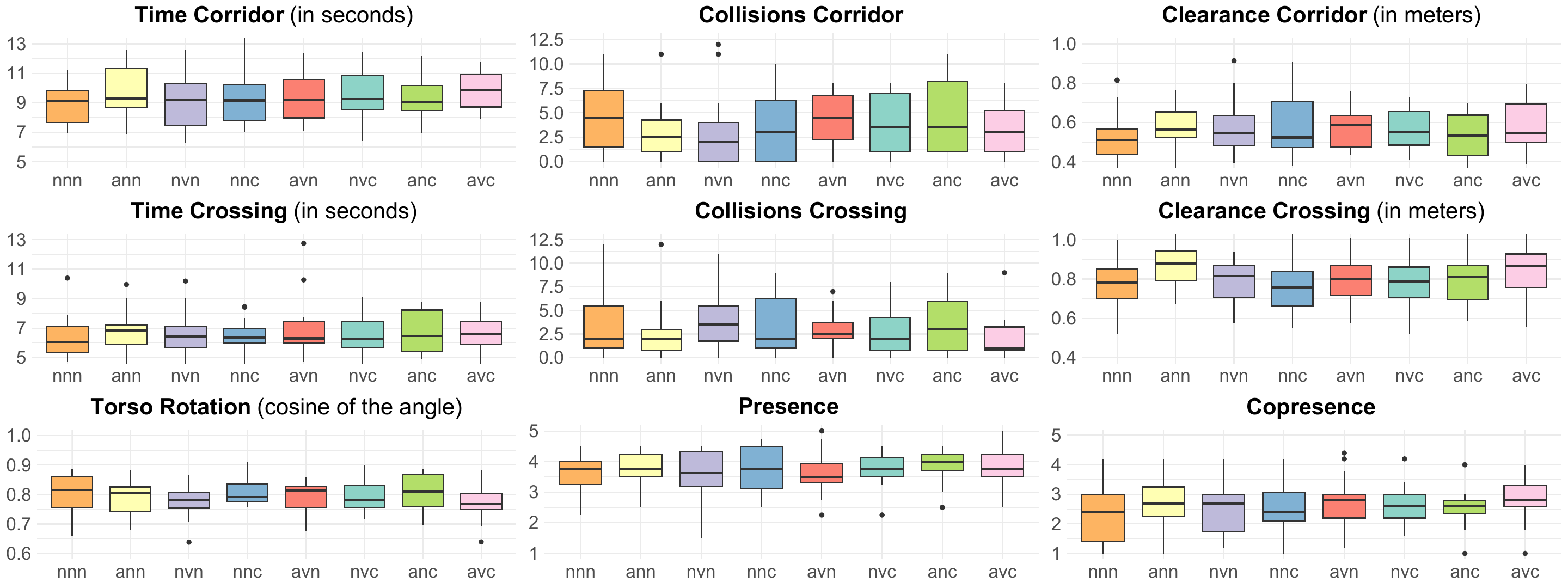}
    \caption{Box plots for all objective and subjective measurements, except \textit{Path}. Each box encloses the middle 50\% of the data for each condition. The thick horizontal lines denote the medians. \textit{a} stands for enabled \textit{AUDIO}, \textit{v} for \textit{VIBR}, \textit{c} for \textit{COLBLF}. \textit{n} means the corresponding factor is disabled. }
    \label{fig:obj_dependent_vars}
\end{figure*}

In this section, we present an overview of the results obtained from the statistical analysis. Figure~\ref{fig:obj_dependent_vars} shows the results of the metrics detailed in Section~\ref{sec:design:measures}. For an in-depth discussion and interpretation of these findings, please refer to Section~\ref{sec:discussion}. 

Shapiro-Wilk tests indicated significant deviations from normality in some instances. As a result, all analyses are carried out using non-parametric tests.

Additionally, we compare the differences between the pretest and all-factors-disabled conditions for all dependent variables. We employ two-sided Wilcoxon tests to compare between groups, and in the case of the dependent variable \textit{Path}, we use a binomial generalized linear mixed-effects model. In all cases, we accept the null hypothesis. Hence, it seems that participants were not affected by sensitization effects such as learning, after effects of main factors (e.g., \textit{VIBR}), or any other effect due to the repeated tasks. 

\subsection{Objective metrics}
\label{sec:objective_metrics}

\subsubsection{Task metrics: Collisions, Time, Clearance} 
To examine the within-subjects factors \textit{AUDIO}, \textit{VIBR}, \textit{COLBLF}, we conducted a repeated measures ART analysis of variance (ANOVA). \added{We report effect sizes using partial eta-squared ($\eta^2_p$). Following Cohen's guidelines \cite{cohen1988}, these can be interpreted as follows: large~($\mathord{>}0.14$), medium~($\mathord{>}0.06$) and small~($\mathord{>}0.01$).} The results indicated that users completed the Corridor scenario significantly slower (higher $\mathit{Time_{corridor}}$) when \textit{AUDIO} was enabled ($F\added{_{(1, 105)}} = 4.57$, $p = 0.035$, $\eta^2_p = 0.04$). However, no significant effect was found for \textit{VIBR} and \textit{COLBLF} on these metrics. Additionally, the interaction between \textit{VIBR} and \textit{COLBLF} was significant ($F\added{_{(1, 105)}} = 5.58$, $p = 0.020$, $\eta^2_p = 0.05$). When both factors were enabled, users completed the Corridor scenario more slowly. Similarly, considering $\mathit{Clearance_{crossing}}$, we found that users significantly increased clearance when \textit{AUDIO} was enabled ($F\added{_{(1, 105)}} = 6.27$, $p = 0.014$, $\eta^2_p = 0.06$; \added{without $10\,\%$ cutoff: $F_{(1, 105)} = 4.46$, $p = 0.036$, $\eta^2_p = 0.04$}). We did not find significant differences for the other dependent variables. 

\subsubsection{Users' motion: Torso rotation} 
We conducted the same tests for the motion-captured data variable \textit{Torso}. In this case, we only found a significant main effect for \textit{VIBR} ($F\added{_{(1, 105)}} = 6.28$, $p = 0.014$, $\eta^2_p = 0.06$; \added{without $10\,\%$ cutoff: $F_{(1, 105)} = 5.90$, $p = 0.016$, $\eta^2_p = 0.05$}). Torso rotations were significantly higher when \textit{VIBR} was enabled.

\subsubsection{Path choice}
Regarding the binary \textit{Path} dependent variable, we conducted a binomial generalized linear mixed-effects model for within-subjects factors. For a complete understanding of the model and its interpretation, we report the results in Figure~\ref{fig:path_probability} and Table~\ref{tab:binomial_path}.
The percentage of choices per condition can be seen in Figure~\ref{fig:path_choice_percentage}. \textit{AUDIO} and \textit{COLBLF} influenced participants to choose the detour path in the corridor scenario. The interaction between \textit{AUDIO} and \textit{COLBLF}, i.e., when both factors were enabled, negatively affected the variable \textit{Path}, and thus, influenced users to choose the direct path. Contrarily, the interaction between all factors positively affected \textit{Path}.

\subsection{Subjective responses}

\subsubsection{Presence and Copresence}
We also report the results when considering the questionnaire answers. As before, we conducted a repeated measures ART ANOVA, using \textit{Presence} and \textit{Copresence} as dependent variables. For \textit{Presence}, we only found a significant main effect for \textit{COLBLF} ($F\added{_{(1, 105)}} = 6.63$, $p = 0.011$, $\eta^2_p = 0.06$). Regarding \textit{Copresence}, results reported a significant main effect for \textit{AUDIO} ($F\added{_{(1, 105)}} = 9.51$, $p = 0.002$, $\eta^2_p = 0.08$). Therefore, \textit{COLBLF} increased the sense of presence, and \textit{AUDIO} that of copresence.

\subsubsection{Open-ended questions}
We now examine open-ended questions posed to participants upon the conclusion of the study. To assess the sentiment behind these responses, we employ sentiment analysis techniques \cite{liu:2012, pang2008opinion} as employed in recent virtual reality studies \cite{slater:2023}. We used the \emph{sentimentr} R package \cite{rinker:2021} to analyze the positivity of responses for each individual question. This package considers valence shifters, enabling us to assess positivity at the sentence level and subsequently average these values to obtain an overall sentiment score for each answer. We report the number of words analyzed $n$, the standard deviation $\sigma$, and the average sentiment $\mathrm{\bar{x}}$ in Table~\ref{tab:open_questions}. Our analysis reveals an overall positive sentiment for questions \textbf{Q1} ($\mathrm{\bar{x}} = 0.09$), \textbf{Q2} ($\mathrm{\bar{x}} = 0.09$) and \textbf{Q4} ($\mathrm{\bar{x}} = 0.21$). Conversely, negative sentiment was observed for question \textbf{Q3} ($\mathrm{\bar{x}} = -0.09$). These results indicate that participants generally had a positive experience and found the auditory feedback to be positive. Notably, participants largely agreed with the statements in \textbf{Q4}, yielding a high average sentiment score. 
For \textbf{Q4}, which targeted a more specific question, we further conducted manual classification of the responses to discern whether participants agreed or disagreed with the statement. Consistent with the high sentiment score, $81\%$ of participants answered affirmatively to \textbf{Q4}.

\section{Discussion}
\label{sec:discussion}



In this section, we revisit the original research questions of the study and analyze the results to summarize the main findings.

\subsection{People response to virtual crowds}
\begin{figure}[t]
    \centering
    \includegraphics[width=1.0\linewidth, alt={Percentage of time participants walked using detour and direct paths for each condition. }]{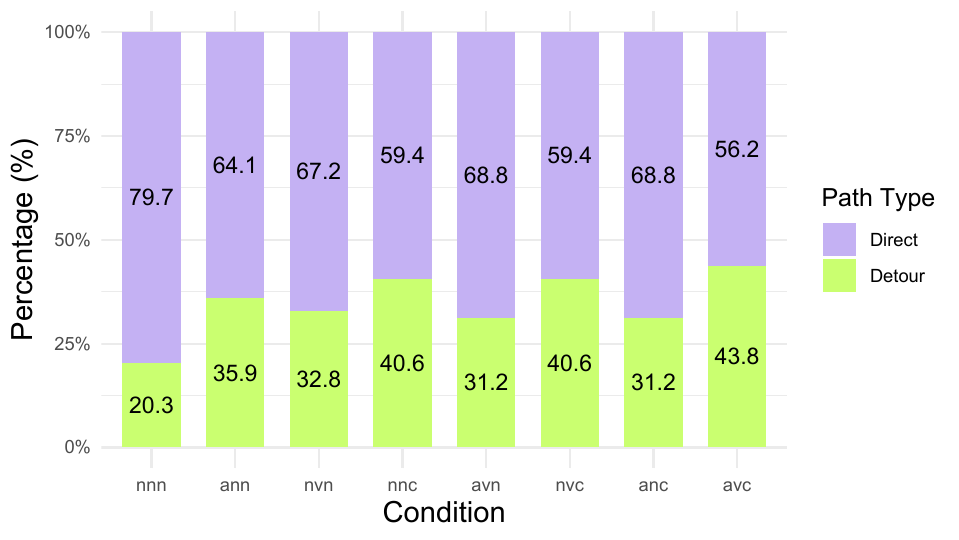}
    \vspace*{-6mm}
    \caption{Percentage of time participants walked, in the Corridor scenario, using a detour path (green) and direct path (purple) for each condition.}
    \label{fig:path_choice_percentage}
\end{figure}

Expected collision feedback plays an important role in shaping participants' behavior towards virtual crowds. As detailed in Section~\ref{sec:objective_metrics}, in the absence of expected collision feedback, \added{it seems that} people tend to walk closer to the virtual \replaced{crowds}{crows} and even pass through them. It becomes particularly evident in situations where participants are presented with both \textit{Direct} and \textit{Detour} pathways. When any form of collision feedback\textemdash \textit{AUDIO} \added{(odds ratio = 2.60)}, \textit{VIBR} \added{(odds ratio = 2.19)}, or \textit{COLBLF} \added{(odds ratio = 3.32)}\textemdash is introduced, participants demonstrate a tendency to opt for the detour yet safer route (see Figure~\ref{fig:path_choice_percentage}). In particular, we found \textit{AUDIO} and \textit{COLBLF} to significantly affect the path decision. Similarly, in the crossing scenario, an increased sense of caution is observed, with individuals maintaining larger \textit{Clearance} from virtual characters. Notably, with \textit{VIBR} enabled, participants rotate their torso to avoid collision more often, or react to the vibrotactile feedback, aligning their actions with findings from prior studies \cite{berton:2022,koilias:2020}. \added{However, it is important to note that in the case of \textit{Clearance} and \textit{Torso}, we observed medium to small effect sizes. These results suggest the need for further studies to fully understand their implications.}


We found no significant effect (for none of the three factors) on the \textit{Collisions} metric, even though individual or combined factors showed noticeable impacts on other measures. This contrasts with a previous study by Berton \textit{et al.} \cite{berton:2022}. A potential reason for this, besides our smaller sample size, could be the dynamics of our experiment: while the prior research used a static crowd with more agents, giving participants enough time to build a strategy, our setup featured moving agents and a relatively smaller number of them, leading to fewer collisions overall.

Beyond behavioral changes, the expectation of collision feedback also \added{appears to} shape participants' subjective perceptions. The presented \textit{Presence} and \textit{Copresence} results underscore this, suggesting participants \added{could} have an increased sense of presence when believing there may be physical collisions with real people and also \added{appear to be} more copresent with the virtual crowds when they heard virtual characters make sounds during collisions. \added{As before, we observed small to medium effect sizes.}

\begin{figure}[t]
\centering
\includegraphics[width=\linewidth,alt={Predicted probability of the detour path from binomial generalized linear mixed-effects models of each factor. }]{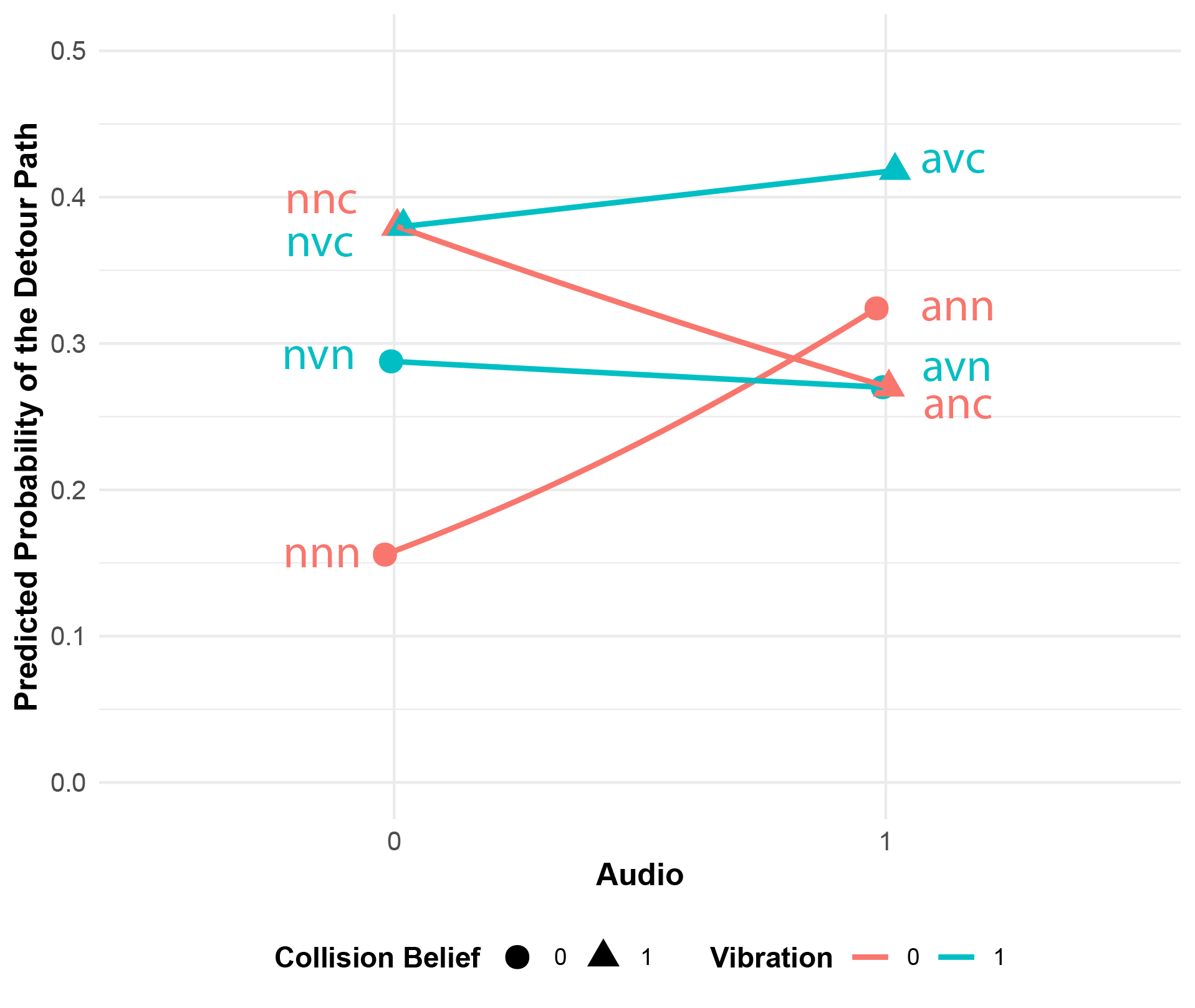}
\vspace*{-6mm}
\caption{Predicted probability of the detour path from binomial generalized linear mixed-effects models of \textit{AUDIO}, \textit{VIBR} and \textit{COLBLF} on \textit{Path} ($0$ stands for factor disabled, $1$ stands for factor enabled).}
\label{fig:path_probability}
\end{figure}

\begin{figure*}
    \centering
    \includegraphics[width=1\linewidth,alt={The trajectories taken by participants in the Corridor task across four trials in two conditions. One is all factors disabled; the other one is only COLBLF enabled. }]{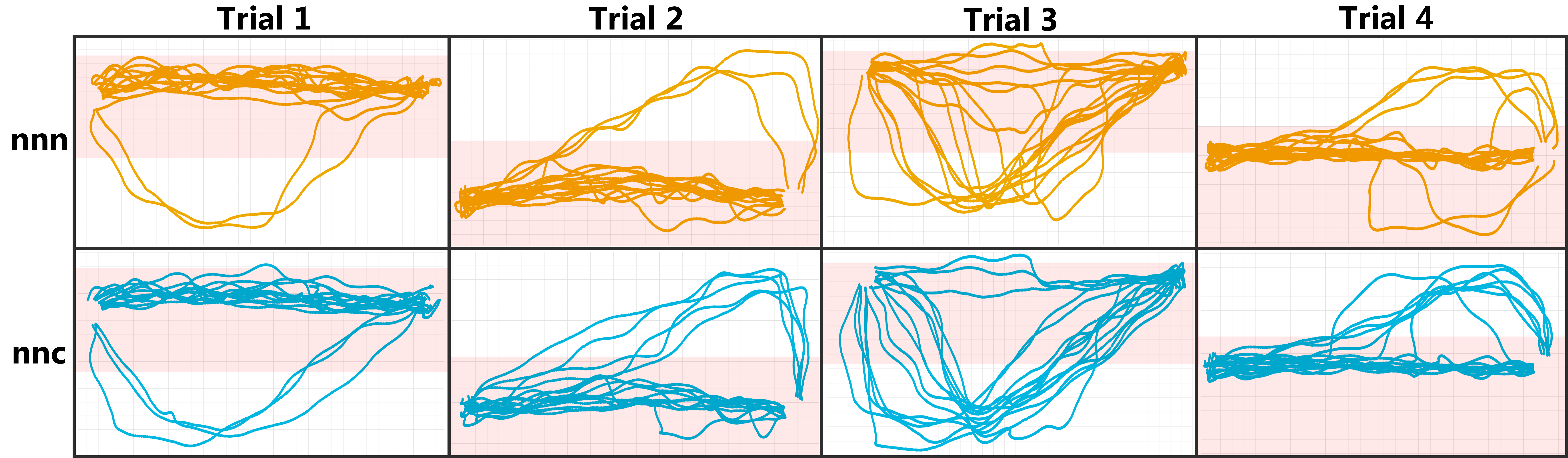}
    \caption{The trajectories taken by participants in the Corridor task across four trials are depicted, with the first row representing the condition where all factors are disabled and the second row highlighting the condition with only \textit{COLBLF} enabled. Areas marked in red indicate high agent density, thus, the area where the \textit{Direct Path} appears, whereas the white area indicates where the \textit{Detour Path} appears. Notably, when \textit{COLBLF} is activated, participants generally opt for detour paths to navigate through the crowd. Specifically, in the third trial, a visible empty space can be seen in the middle of the corridor in the second row, suggesting that participants actively avoided this area. In the fourth trial, in the condition without \textit{COLBLF}, some participants chose to navigate toward the bottom of the corridor where the agent density is highest, a behavior that is absent when \textit{COLBLF} is enabled.}
    \label{fig:paths_nnn_nnc}
\end{figure*}

\begin{table}
\centering
\begin{tabu}{lrccc}
\toprule
Fixed Effect & Estimate & Std. Error & Odds R. & \( p \) \\
\midrule
\textbf{\textit{(Intercept)}} & -1.69 & 0.44 & 0.18 & \textbf{\textless\,0.001} \\
\textbf{\textit{AUDIO}} & 0.95 & 0.44 & 2.60 & \textbf{0.031} \\
\textit{VIBR} & 0.78 & 0.45 & 2.19 & 0.08 \\
\textbf{\textit{COLBLF}} & 1.20 & 0.44 & 3.32 & \textbf{0.006} \\
\textit{A:V} & -1.04 & 0.61 & 0.35 & 0.09 \\
\textbf{\textit{A:C}} & -1.46 & 0.60 & 0.23 & \textbf{0.015} \\
\textit{V:C} & -0.78 & 0.60 & 0.46 & 0.19 \\
\textbf{\textit{A:V:C}} & 1.71 & 0.84 & 5.51 & \textbf{0.041} \\
\bottomrule \addlinespace[0.05cm]
\end{tabu}
\caption{Binomial generalized linear mixed-effects models of \textit{AUDIO}, \textit{VIBR} and \textit{COLBLF} on \textit{Path}. Estimate is the coefficient for the predictor in the logistic model. Std. Error is the standard error of the coefficient estimate. Odds R. represents the odds of the user choosing the detour path. Values are reported for enabled fixed effects, e.g., Odds R. in \textit{AUDIO} refers to the probability of a user choosing the detour path when \textit{AUDIO} is on.}
\label{tab:binomial_path}
\end{table}

\subsection{Auditory and vibrotactile collision feedback}
In Section~\ref{sec:objective_metrics}, the results highlighted the distinct ways in which the \textit{AUDIO} and \textit{VIBR} factors shaped participants' avoidance behaviors. 

The \textit{AUDIO} factor significantly influenced \emph{global} behaviors, which refer to high-level strategies participants use when walking towards the target. This was evidenced by participants' \added{small to medium} tendency to maintain larger distances from crowd agents and an increased task completion time.

Conversely, the \textit{VIBR} factor \replaced{appears to have}{prominently} affected \emph{local} behaviors, i.e., immediate and specific reactions to nearby agents. For instance, the Torso metric suggests that participants rotated their torso to narrowly avoid nearby virtual characters, indicating that vibrotactile feedback \added{may have} played a role in their immediate, moment-to-moment decisions. Such findings align with previous work, which found that haptic feedback only caused people to rotate their torso more often but did not affect their \emph{global} behaviors \cite{berton:2022}. 

When participants provided feedback on their experiences, auditory cues received a generally positive response. We believe that voices like ``Watch out!'' and ``Ouch'' raised empathy towards virtual characters, prompting participants to move through virtual crowds more carefully instead of ignoring their presence. Conversely, vibrotactile feedback received more negative opinions. Some participants mentioned that the vibrations felt intrusive or startling, detracting from the immersive experience. Another point to ponder is the distinction between the haptic sensations produced by the armbands and the genuine tactile experience one might encounter during real-world collisions.



\subsection{Perceived risk of collision}

In the Corridor scenario, the \textit{COLBLF} factor notably influenced participants towards choosing the detour path when presented with a group of agents offering both detour and direct path options. Additionally, this factor resulted in a significantly higher \textit{Presence} score. This suggests that the presence feeling \replaced{may have been}{was} intensified when participants believed real people existed within the virtual crowd. This \replaced{may}{can} be explained because participants increased their focus on the task of avoiding virtual agents. Participants could only avoid the potential real human(s) by avoiding the virtual avatars. This sentiment was further corroborated in the positive post-experiment responses to question \textbf{Q4}. However, as noted in 
Figure~\ref{fig:path_probability}
, when both \textit{COLBLF} and \textit{AUDIO} are enabled, the detour path choice probability decreases when compared to having only \textit{COLBLF} enabled. Our best explanation for this is that when participants started listening to the virtual avatars, they may have realized no real person was walking in the crowd, thus breaking the collision belief. This does not happen when \textit{VIBR} is enabled, significantly increasing the time to finish the corridor scenario. We believe having vibrotactile feedback further reinforces collision belief, as such vibration is somehow close to a gentle collision or graze with a real human. \added{Nonetheless, this perspective is opposed to the negative feedback from participants concerning the plausibility of the vibrations. Thus, more research is needed to investigate this issue.} \added{In addition, while these interactions are statistically significant, it is important to note that the associated effect sizes are relatively small.}

Interestingly, while the \textit{COLBLF} factor \replaced{appears to have}{clearly} influenced the global behavior of the participants and how immersed they felt, it did not seem to change their local behavior. This suggests that while participants might change their overall approach due to \textit{COLBLF}, their natural reactions when moving through crowds remained consistent, regardless of whether they believed they might bump into real people in the virtual space.



\section{Conclusions and Future Work}
\label{sec:conclusions}

In this paper, we have explored the effect of collision feedback expectancy on VR users moving through virtual crowds. For several reasons, we have considered explicitly the type of feedback the user expects vs. the type of feedback the user gets. First, this allowed us to compare against a novel factor (\textit{COLBLF}) that is mostly based on what the users believe may happen (a collision with a real person embodied in one of the avatars) instead of the true feedback they will actually receive. Second, because actual collision feedback is always approximate: even if the full geometry of the participant avatar and the other avatars is taken into consideration to detect collisions (which is not typically the case for performance reasons), it is very unlikely that the self-avatar faithfully reproduces the shape and motion of the VR user, even if wearing a MoCap suit. This means that subtle collisions will not be detected by the system, and thus no collision feedback will be produced, no matter the condition (\textit{AUDIO}, \textit{VIBR}, \textit{COLBLF}). In other words, although \textit{COLBLF} provides no actual feedback during the experiment (unlike \textit{AUDIO} and \textit{VIBR}), in practice such a difference might go unnoticed unless the user makes absolutely no effort to avoid large collisions. 

The main finding of our study is that the expected collision feedback significantly influences both the behavior of the participants and their subjective perceptions of presence and copresence. Our best explanation for the results distinguishes between metrics referring to global behavior (time, clearance, path) vs. local behavior (torso rotation).  The introduction of a perceived risk of actual collision seems to significantly affect global behavior and increase the sense of presence. Auditory feedback, in the form of clearly audible complaints, appears to have a similar effect on global behavior and additionally increases the sense of copresence. In contrast, vibrotactile feedback was primarily effective in influencing local behavior, connected to immediate decisions. 

We hope that these findings will help researchers and developers both in designing suitable VR environments involving virtual crowds, as well as anticipating the effect of collision feedback strategies on user behavior and experience. 


\textbf{Limitations: }
Since our sample size (N=16) was relatively small, we might have missed some significant effects. \added{In addition, it should be noted that some statistically significant findings exhibit a small to moderate effect size.} Notice also that the participants of our study were young adults, so some of our findings may not apply to the general population. As with many other user studies in VR, our results are closely dependent on the chosen scenarios, tasks, crowd behavior, as well as the cultural context of the participants. For example, significantly increasing crowd density and/or agent speed might considerably increase the physical effort to avoid collisions, and this might cause some users to decide to disregard collisions, especially if only audio or vibrotactile feedback is expected. Other aspects, such as the cognitive load of the task, might also influence the results. In our experiment, the task required a minimal cognitive load, since users just had to walk towards a clearly visible target. We guess that more complex scenarios involving, for example, search and/or wayfinding tasks could potentially lessen the effect of collision feedback expectancy.

An interesting avenue for future work is actually conducting further experiments to study which of our results apply to scenarios and tasks significantly departing from ours. For example, we are interested in knowing for how long the different collision feedback expectancy conditions have an effect on the measured variables, as we suspect that
tasks with longer completion times might require periodic reinforcements of the expected collision feedback (as in the adaptation scenario), for example in the case of \textit{COLBLF}. We also plan to explore more general strategies for inducing collision belief; although in some settings \textit{COLBLF} is easy and even natural to adopt (for example, in collaborative VR applications with a mixture of human and NPC avatars), since the adaptation phase involves another person, it does not apply for example to VR applications used at home.

\acknowledgments{
This work has received funding from the European Union’s Horizon 2020 research and innovation programme under the Marie Skłodowska-Curie grant agreement No 860768 (CLIPE project), 
and from MCIN/AEI/10.13039/501100011033/FEDER, UE (PID2021-122136OB-C21), "A way to make Europe". Jose Luis Ponton was also funded by the Spanish Ministry of Universities (FPU21/01927).
}

\bibliographystyle{abbrv-doi}

\bibliography{references}
\end{document}